\documentclass[11pt,a4paper]{article}

\usepackage{fullpage}
\usepackage{dcolumn}
\usepackage{amsmath}
\usepackage{amssymb}
\usepackage{mathrsfs}
\usepackage{graphicx}
\usepackage{psfrag}
\usepackage[all,poly,curve,knot,cmtip]{xy}
\usepackage{citesort}

\newcommand {\be} {\begin{eqnarray*}}
\newcommand {\ee} {\end{eqnarray*}}
\newcommand {\bea} {\begin{eqnarray}}
\newcommand {\eea} {\end{eqnarray}}

\newcommand{\bm}[1]{\boldsymbol{#1}}

\newcommand{\fdiff}[2]{\frac{\delta{#1}}{\delta{#2}}}

\newcommand{\tq}{\tilde{q}}

\begin{document}

\title{Topological deformation of isolated horizons}
\author{\textbf{Tom$\acute{\mbox{a}}\check{\mbox{s}}$
Liko}\footnote{Electronic mail: tliko@math.mun.ca}\\
\\{\small \it Department of Physics and Physical Oceanography}\\
{\small \it Memorial University of Newfoundland}\\
{\small \it St. John's, Newfoundland, Canada, A1B 3X7}}


\maketitle

\begin{abstract}

We show that the Gauss-Bonnet term can have physical effects
in four dimensions.  Specifically, the entropy of a black
hole acquires a correction term that is proportional to the
Euler characteristic of the cross sections of the horizon.
While this term is constant for a single black hole, it will
be a non-trivial function for a system with dynamical
topologies such as black-hole mergers: it is shown that for
certain values of the Gauss-Bonnet parameter, the second law
of black-hole mechanics can be violated.

\end{abstract}

\hspace{0.35cm}{\small \textbf{PACS}: 04.70.Bw}




The study of black-hole thermodynamics continues to be one of the most
exciting areas in gravitational theory.  The celebrated four laws of
black-hole mechanics \cite{bekenstein1,bekenstein2,bch} have revealed
a very deep and profound relationship between classical and quantum
aspects of gravitational phenomena.  Among these, the first law relates
the small changes of energy to small changes of surface area of nearby
equilibrium states of a black hole within the phase space of solutions.
This leads to an identification of a multiple of the surface gravity
$\kappa$ on the horizon with the temperature $\mathcal{T}$ of the hole,
and a multiple of the surface area $A$ with the entropy $\mathcal{S}$.
More precisely, the temperature and entropy are
\cite{bekenstein1,bekenstein2,hawking1}
\bea
\mathcal{T} = \frac{\kappa}{2\pi}
\quad
\mbox{and}
\quad
\mathcal{S} = \frac{A}{4G} \, ,
\label{ts}
\eea
with $G$ the Newton constant.  Remarkably, this expression for the
entropy is independent of other properties of the black hole, such
as the electric (or Yang-Mills) charge or rotation.

A general analysis based on Noether charge methods \cite{wald1,iyewal,jkm}
has revealed that modifications to the Bekenstein-Hawking entropy relation
will only present themselves in cases when gravity is non-minimally coupled
to matter, or when the action for gravity is supplemented with
higher-curvature interactions.  The presence of higher-curvature
interactions is important within the context of string theory; the
Kretchman scalar appears in the low-energy effective action from the
heterotic string theory \cite{chsw}.  Of particular interest is the
Gauss-Bonnet (GB) term, which is the only combination of curvature-squared
interactions for which the effective action is ghost-free \cite{zwiebach}.
The complete action for gravity in $D$ dimensions is then \cite{zwiebach}
\bea
S &=& \frac{1}{2k_{D}}\int_{\mathcal{M}}d^{D}x\sqrt{-g}
      (R - 2\Lambda + \alpha\mathcal{L}_{GB})\nonumber\\
\mathcal{L}_{GB} &=& R^{2} - 4R_{ab}R^{ab} + R_{abcd}R^{abcd} \; .
\label{action1}
\eea
In this expression, $g$ is the determinant of the spacetime metric
tensor $g_{ab}$ ($a,b,\ldots\in\{0,\ldots,D-1\}$), $R_{abcd}$ is the
Riemann curvature tensor, $R_{ab}=R_{\phantom{a}acb}^{c}$ is the Ricci
tensor, $R=g^{ab}R_{ab}$ is the Ricci scalar, $k_{D}=8\pi G_{D}$ with
$G_{D}$ the $D$-dimensional Newton constant is the $D$-dimensional
coupling constant, $\Lambda$ is the cosmological constant, and $\alpha$
is the GB parameter.

A common belief within the literature about the action (\ref{action1})
is that in four dimensions the GB term can be discarded because it is
a topological invariant (the Euler characteristic), and only leads to
non-trivial effects in $D\geq5$ dimensions.  However, variation of
$\mathcal{L}_{GB}$ in $D=4$ dimensions gives a surface term; this can
be discarded locally, but becomes an important contribution if the
manifold has boundaries.  So if we are to believe that the GB term
is significant in $D\geq5$ dimensions, then (for a bounded spacetime)
it should be considered to be significant in $D=4$ dimensions as well.
As we will show, inclusion of the GB term in $D=4$ dimensions has
important implications for black-hole mechanics.

We will elaborate on the above point in a moment, in particular, how
variation of $\mathcal{L}_{GB}$ gives rise to a surface term in four
dimensions.  This will be done in the connection formulation of general
relativity.  However, because we are interested in a manifold with
boundaries, we first introduce the boundary conditions; it will be
shown that first variation leads to a well defined action principle.
We consider a four-dimensional spacetime manifold $\mathcal{M}$ of
topology $R\times M$ with the following properties: (a) $\mathcal{M}$
contains a three-dimensional null surface $\Delta$ as inner boundary
(representing the event horizon); and (b) $\mathcal{M}$ is bounded by
three-dimensional (partial) Cauchy surfaces $M^{\pm}$ that intersect
$\Delta$ in two-surfaces $\mathscr{S}^{\pm}$ and extend to the (arbitrary)
boundary at infinity $\mathscr{B}$.  See Figure $1$.
\begin{figure}[t]
\begin{center}
\psfrag{D}{$\Delta$}
\psfrag{Mp}{$M^{+}$}
\psfrag{Mm}{$M^{-}$}
\psfrag{Mi}{$M$}
\psfrag{Sp}{$\mathscr{S}^{+}$}
\psfrag{Sm}{$\mathscr{S}^{-}$}
\psfrag{S}{$\mathscr{S}$}
\psfrag{B}{$\mathscr{B}$}
\psfrag{M}{$\mathcal{M}$}
\includegraphics[width=3.2in]{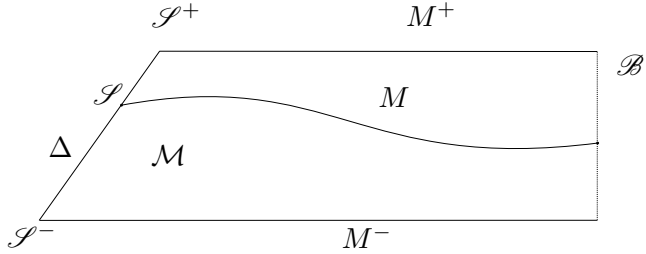}
\caption{The region of the four-dimensional spacetime $\mathcal{M}$
being considered has an internal boundary $\Delta$ representing
the event horizon, and is bounded by two (partial) Cauchy surfaces
$M^{\pm}$ which intersect $\Delta$ in two-surfaces
$\mathscr{S}^{\pm}$ and extend to the boundary at infinity
$\mathscr{B}$.}
\end{center}
\end{figure}

A three-dimensional null hypersurface $\Delta$ (with topology
$R\times\mathscr{S}$) together with a degenerate metric $q_{ab}$
of signature $0++$ and a null normal $\ell_{a}$ is said to be a
non-expanding horizon if: (a) the expansion $\theta_{(\ell)}$ of
$\ell_{a}$ vanishes on $\Delta$; (b) the field equations hold on
$\Delta$; and (c) the matter stress-energy tensor is such that
$-T_{\phantom{a}b}^{a}\ell^{b}$ is a future-directed causal vector.
Condition (a) implies that the rotation tensor is zero.  Condition
(c) is the dominant energy condition imposed on any matter fields
that may be present in the neighbourhood of the horizon.  These
conditions along with the Raychaudhuri equation imply that the shear
tensor also vanishes.  In turn, this implies that
$\nabla_{\!\underleftarrow{a}}\ell_{b}\approx\omega_{a}\ell_{b}$.
(The underarrow indicates pullback to $\Delta\subset\mathcal{M}$;
``$\approx$'' denotes equality restricted to $\Delta$.)  Thus the
one-form $\omega$ is the natural connection (in the normal bundle)
induced on the horizon.

The ``time-independence'' of $\omega$ on $\Delta$ captures the notion
of weak isolation.  That is, a non-expanding horizon together with an
equivalence class of null normals $[\ell]$ becomes a weakly isolated
horizon if $\pounds_{\ell}\omega_{a}=0$ for all $\ell\in[\ell]$ (where
$\ell^{\prime}\sim\ell$ if $\ell^{\prime}=c\ell$ for some constant
$c$).  This condition is a restriction on the rescaling freedom of
$\ell$.  It turns out that this condition is enough to establish
the zeroth law: the surface gravity
$\kappa_{(\ell)}=\ell^{a}\omega_{a}$ is constant over the surface
$\Delta$ of a weakly isolated horizon.  This form of the zeroth law
was first established in \cite{afk}.

Let us now look at the action principle, and the implications of
the boundary conditions on the first variation.  This is most
transparent in the connection formulation of general relativity,
where the action (\ref{action1}) becomes
\bea
S = \frac{1}{2k_{4}}\int_{\mathcal{M}}\Sigma_{IJ} \wedge \Omega^{IJ}
    - 2\Lambda\bm{\epsilon}
    + \alpha\epsilon_{IJKL}\Omega^{IJ} \wedge \Omega^{KL} \; .
\label{action2}
\eea
This action depends on the coframe $e^{I}$ and the connection
$A_{\phantom{a}J}^{I}$.  The coframe determines the metric
$g_{ab}=\eta_{IJ}e_{a}^{\phantom{a}I} \otimes e_{b}^{\phantom{a}J}$,
two-form $\Sigma_{IJ}=(1/2)\epsilon_{IJKL}e^{K} \wedge e^{L}$ and
spacetime volume four-form
$\bm{\epsilon}=e^{0} \wedge e^{1} \wedge e^{2} \wedge e^{3}$, where
$\epsilon_{IJKL}$ is the totally antisymmetric Levi-Civita tensor.  The
connection determines the curvature two-form
\bea
\Omega_{\phantom{a}J}^{I} = dA_{\phantom{a}J}^{I}
+A_{\phantom{a}K}^{I} \wedge A_{\phantom{a}J}^{K}
= \frac{1}{2}R_{\phantom{a}JKL}^{I}e^{K} \wedge e^{L} \, ,
\eea
with $R_{\phantom{a}JKL}^{I}$ as the Riemann tensor.  Internal indices
$I,J,\ldots\in\{0,\ldots,3\}$ are raised and lowered using the Minkowski
metric $\eta_{IJ}=\mbox{diag}(-1,1,1,1)$.  The gauge covariant derivative
$\mathscr{D}$ acts on generic fields $\Psi_{IJ}$ such that
\bea
\mathscr{D}\Psi_{\phantom{a}J}^{I}
= d\Psi_{\phantom{a}J}^{I}
  + A_{\phantom{a}K}^{I} \wedge \Psi_{\phantom{a}J}^{K}
  - A_{\phantom{a}J}^{K} \wedge \Psi_{\phantom{a}K}^{I} \; .
\eea
In general, the equations of motion are given by $\delta S=0$, where $\delta$
is the first variation; i.e. the stationary points of the action.  In the
present case, the equations of motion are obtained from independently varying
the action (\ref{action2}) with respect to the coframe and connection.  Denoting
the pair $(e,A)$ collectively as a generic field variable $\Psi$, the first
variation gives
\bea
\delta S = \frac{1}{2k_{4}}\int_{\mathcal{M}}E[\Psi]\delta\Psi
           - \frac{1}{2k_{4}}\int_{\partial\mathcal{M}}J[\Psi,\delta\Psi] \; .
\label{first}
\eea
Here $E[\Psi]=0$ symbolically denotes the equations of motion.  Specifically,
these are:
\bea
\fdiff{S}{A} &\rightarrow&
\mathscr{D}(\Sigma_{IJ} + 2\alpha\epsilon_{IJKL}\Omega^{KL}) = 0
\label{em1}\\
\fdiff{S}{e} &\rightarrow&
\epsilon_{IJKL}e^{J} \wedge (\Omega^{KL} - 2\Lambda e^{K} \wedge e^{L}) = 0 \; .
\label{em2}
\eea
The first of these reduces to $\mathscr{D}e=0$ by virtue of the Bianchi
identity.  The surface term $J$ is given by
\bea
J[\Psi,\delta\Psi] &=& \widetilde{\Sigma}_{IJ} \wedge \delta A^{IJ},\nonumber\\
\widetilde{\Sigma}_{IJ} &\equiv& \Sigma_{IJ} + 2\alpha\epsilon_{IJKL}\Omega^{KL} \; .
\label{surface}
\eea
If the integral of $J$ on the boundary $\partial\mathcal{M}$ vanishes then the
action principle is said to be differentiable.  We must show that this is the
case.  Because the fields are held fixed at $M^{\pm}$ and at $\mathscr{B}$, $J$
vanishes there.  So we only need to show that $J$ vanishes at the inner boundary
$\Delta$.  To show that this is true we need to find an expression for $J$ in
terms of $A$ and $\widetilde{\Sigma}$ pulled back to $\Delta$.  This is
accomplished by fixing an internal Newman-Penrose basis consisting of the null
vectors $(\ell,n,m,\bar{m})$ such that $\ell=e_{0}$, $n=e_{1}$,
$m=(e_{2}+ie_{3})/\sqrt{2}$, and $\bar{m}=(e_{2} - ie_{3})/\sqrt{2}$;
normalizations are such that $\ell \cdot n = -1$, $m \cdot \bar{m} = 1$, and
all other contractions are zero.  The pullback of $A$ can be expressed as
\bea
A_{\underleftarrow{a}IJ}
\approx
- 2\ell_{\left[I\right.}n_{\left.J\right]}\omega_{a}
+ X_{a}\ell_{\left[I\right.}m_{\left. J\right]}
+ Y_{a}\ell_{\left[I\right.}\bar{m}_{\left. J\right]}
+ Z_{a}m_{\left[I\right.}\bar{m}_{\left. J\right]} \, ,
\label{pullbackofa}
\eea
where $X_{a}$, $Y_{a}$ and $Z_{a}$ are one-forms in the cotangent bundle
$T^{*}(\Delta)$.  It follows that the variation of (\ref{pullbackofa}) is
\bea
\delta A_{\underleftarrow{a}IJ}
\approx
- 2\ell_{\left[I\right.}n_{\left.J\right]}\delta\omega_{a}
+ \delta X_{a}\ell_{\left[I\right.}m_{\left. J\right]}
+ \delta Y_{a}\ell_{\left[I\right.}\bar{m}_{\left. J\right]}
+ \delta Z_{a}m_{\left[I\right.}\bar{m}_{\left. J\right]} \; .
\label{variationofpullbackofa}
\eea
To find the pullback to $\Delta$ of $\widetilde{\Sigma}$, we use the
decompositions
\bea
e_{a}^{\phantom{a}I}
&=& -\ell^{I}n_{a} - n^{I}\ell_{a}
    + m^{I}\bar{m}_{a} + \bar{m}^{I}m_{a}\\
\epsilon_{IJKL}
&=& i\ell_{I} \wedge n_{J} \wedge m_{K} \wedge \bar{m}_{L} \; .
\eea
The pullback of $\Sigma$ is \cite{afk}
\bea
\underleftarrow{\Sigma}_{IJ}
\approx 2\ell_{\left[I\right.}n_{\left. J\right]}\bm{\tilde{\epsilon}}
          + 2n \wedge (im\ell_{\left[I\right.}\bar{m}_{\left. J\right]}
          - i\bar{m}\ell_{\left[I\right.}m_{\left. J\right]}) \; .
\label{pullbackofsigmaij}
\eea
Here we have defined the area form $\bm{\tilde{\epsilon}} = im \wedge \bar{m}$.
To calculate the pullback of the curvature we use the definition
\bea
\Omega_{abIJ} = R_{IJKL}e_{\left[a\right.}^{\phantom{a}K}e_{\left. b\right]}^{\phantom{a}L} \, ,
\label{curvature}
\eea
whence
\bea
\underleftarrow{\Omega}_{abIJ}
\approx
2R_{IJKL}\left[\ell^{K}m^{L}(\bar{m} \wedge n)
               + \ell^{K}\bar{m}^{L}(m \wedge n)\right.
               + \left.m^{K}\bar{m}^{L}(\bar{m} \wedge m)\right] \; .
\label{pullbackofcurvature}
\eea
Now, we note that
$\underleftarrow{\Sigma}_{IJ} \wedge \delta\underleftarrow{A}^{IJ}\approx
2\bm{\tilde{\epsilon}} \wedge \delta\omega$.  Using this together with the
expressions (\ref{variationofpullbackofa}), (\ref{pullbackofsigmaij}) and
(\ref{curvature}), we find that the surface term (\ref{surface}) becomes
\bea
J[\Psi,\delta\Psi]
&\approx& \left[\bm{\tilde{\epsilon}} + 2i\alpha R_{IJKL}
          m^{I}\bar{m}^{J}e^{K} \wedge e^{L}\right]
          \wedge \delta \omega\nonumber\\
& &
          - \frac{i\alpha}{2} R_{IJKL}\ell^{I}\left[m^{J}\delta X\right.
          + \bar{m}^{J}\delta Y
          - \left. n^{J}\delta Z\right] \wedge e^{K} \wedge e^{L} \; .
\label{pullbackofcurrent}
\eea
(A factor of $2$ has been absorbed into the coefficient outside the integral
in (\ref{first}).)  For an isolated horizon, the Riemann tensor is severely
restricted.  This results in considerable simplification of
(\ref{pullbackofcurrent}).  Details of these simplifications are worked out
in the appendix in \cite{likboo} for multi-dimensional weakly isolated and
non-rotating horizons; here we just state the results and refer the reader
to that article for more details.  In particular, the pullback to $\Delta$
of the Riemann tensor is equivalent to the Riemann tensor $\mathcal{R}_{IJKL}$
of the two-dimensional cross sections of $\Delta$.  That is,
\bea
\tq_{a}^{\phantom{a}e}\tq_{b}^{\phantom{a}f}\tq_{c}^{\phantom{a}g}
\tq_{d}^{\phantom{a}h}R_{efgh} = \mathcal{R}_{abcd} \; .
\eea
The $\tq$ in this expression is the projection tensor onto $\mathscr{S}$
defined by $\tq_{a}^{\phantom{a}b}=q_{a}^{\phantom{a}b}+\ell_{a}n^{b}$.
Further simplification occurs if the horizon is non-rotating, in which case
we have that $\omega_{a}=-\kappa_{(\ell)}n_{a}$.  Using this with the fact
that the expansion, rotation and shear are all zero on $\Delta$ implies that
$R_{\underleftarrow{ab}\phantom{a}d}^{\phantom{aa}c}\ell^{d} = 0$; with these
considerations, it turns out that the only non-vanishing contribution in
(\ref{pullbackofcurrent}) is
$\mathcal{R}_{IJKL}m^{I}\bar{m}^{J}m^{K}\bar{m}^{L}\approx\mathcal{R}$,
with $\mathcal{R}$ the Ricci scalar of the cross sections $\mathscr{S}$
of the horizon.  Hence the current (\ref{pullbackofcurrent}) becomes
\bea
J[\Psi,\delta\Psi]
\approx \bm{\tilde{\epsilon}}(1 + 2\alpha\mathcal{R}) \wedge \delta \omega \; .
\label{simplifiedpullbackofcurrent}
\eea
The final step is to note that $\delta\ell\propto\ell$ for some $\ell$ fixed in
$[\ell]$, and this together with $\pounds_{\ell}\omega=0$ implies that
$\pounds_{\ell}\delta\omega=0$.  However, $\omega$ is held fixed on $M^{\pm}$ which
means that $\delta\omega=0$ on the initial and final cross sections of $\Delta$
(i.e. on $M^{-}\cap\Delta$ and on $M^{+}\cap\Delta$), and because $\delta\omega$ is
Lie dragged on $\Delta$ it follows that $J\approx0$.  Therefore the surface term
$J|_{\partial\mathcal{M}}=0$ for four-dimensional gravity with GB term, and we
conclude that the equations of motion $E[\Psi]=0$ follow from the action principle
$\delta S=0$.

The expression (\ref{simplifiedpullbackofcurrent}) for the current pulled back to
$\Delta$ is the same as the one that was obtained for a multidimensional horizon
\cite{likboo}.  The calculation presented in this paper may seem like a simple
re-calculation of $J$ that was presented in \cite{likboo}, with the dimensionality
restricted to $D=4$.  However, we believe that the calculation presented here is
a necessary one because the phase space of the horizon in four dimensions differs
from the phase space of the corresponding horizon in $D\geq5$ dimensions.
Specifically, the GB density in four dimensions is
$\mathcal{L}_{GB} \sim \epsilon_{IJKL}\Omega^{IJ} \wedge \Omega^{KL}$ which only
depends on the connection.  In $D\geq5$ dimensions the GB density becomes
$\mathcal{L}_{GB} \sim \Sigma_{IJKL} \wedge \Omega^{IJ} \wedge \Omega^{KL}$, with
$\Sigma$ defined by
$\Sigma_{I_{1}\ldots I_{m}}=\epsilon_{I_{1}\ldots I_{m}I_{m+1}\ldots I_{D}}e^{I_{m+1}}
\wedge \cdots \wedge e^{I_{D}}$.  In addition to the connection, this term also
depends on the coframe through $\Sigma$.  As a result, the equations of motion are
more complicated and physically different from their four-dimensional counterparts.
Among other consequences, the equation of motion for the connection does not
constrain the torsion two-form to vanish in higher dimensions.


The calculation of the first law from the surface term is now essentially
the same as in \cite{likboo}.  For an appropriate normalization of some
time evolution vector field $t$ that points in the direction of $\ell$,
and defining the surface gravity $\kappa_{(t)}=t\cdot\omega$,
the first law for the horizon with energy $\mathcal{E}_{\Delta}$ is
\bea
\delta \mathcal{E}_{\Delta} = \frac{\kappa_{(t)}}{k_{4}}\delta\oint_{\mathscr{S}}
                    \bm{\tilde{\epsilon}}(1 + 2\alpha\mathcal{R}) \; .
\label{firstlaw}
\eea
In its standard form, the first law of thermodynamics (for a quasi-static
process) is
$\delta \mathcal{E}=\mathcal{T}\delta \mathcal{S}+(\mbox{work terms})$.
Here, the temperature is $\mathcal{T}=\kappa_{(t)}/2\pi$, whence the
entropy of the horizon is
\bea
\mathcal{S} = \frac{1}{4G}\oint_{\mathscr{S}}\bm{\tilde{\epsilon}}(1 + 2\alpha\mathcal{R}) \; .
\label{entropy3}
\eea
This differs from the Bekenstein-Hawking expression (\ref{ts}).  Therefore,
the GB term gives rise to a correction even though it is a topological invariant
of the manifold and does not show up in the equations of motion.  This happens
because the GB term contributes a surface term which cannot be discarded in the
covariant phase space.

Here, the spaces $\mathscr{S}$ are two-dimensional: \emph{the correction term is
(a multiple of) the Euler characteristic $\chi(\mathscr{S})$ of the cross sections
of the horizon}.  This is consistent with the conclusions in \cite{jacmye}, but
much more general because we did not specify any properties of the space
$\mathscr{S}$.  By the GB theorem, we have that
$\oint_{\mathscr{S}}\bm{\tilde{\epsilon}}\mathcal{R}=4\pi\chi(\mathscr{S})$.  The
entropy (\ref{entropy3}) is therefore
\bea
\mathcal{S} = \frac{1}{4G}\left[A + 8\pi\alpha\chi(\mathscr{S})\right] \; .
\label{entropy4}
\eea
For example, if $\Lambda$ is zero, then by Hawking's topology theorem $\mathscr{S}$
has to be a sphere \cite{hawking2}.  In this case $\chi(\mathscr{S})=2$ and the
entropy becomes $\mathcal{S}=(A + 16\pi\alpha)/(4G)$.  If the cosmological constant
is negative, then physical black holes can have spherical, flat, or even toroidal as
well as higher-genus horizon topologies \cite{blp}.  For a torus,
$\chi(\mathscr{S})=0$ and the Bekenstein-Hawking entropy $\mathcal{S}=A/(4G)$ is
recovered.


For a single black hole, the correction is a constant.  However, this will not
be the case for a system with dynamical topologies such as black-hole mergers
\cite{witt}.  This is a form of topology change, which for a space with a
degenerate metric is unavoidable even in classical general relativity
\cite{horowitz}.  As an example, let us consider the merging of two black holes
-- one with mass $m_{1}$ and entropy
$\mathcal{S}_{1}=[A_{1}+8\pi\alpha\chi(\mathscr{S}_{1})]/4G$, the other with mass
$m_{2}$ and entropy $\mathcal{S}_{2}=[A_{2}+8\pi\alpha\chi(\mathscr{S}_{2})]/4G$.
Before the black holes merge, the total entropy is
\bea
\mathcal{S} &=& \mathcal{S}_{1} + \mathcal{S}_{2}\nonumber\\
            &=& \frac{1}{4G}[A_{1} + A_{2} + 8\pi\alpha(\chi(\mathscr{S}_{1})
                + \chi(\mathscr{S}_{2}))] \; .
\label{entropybefore}
\eea
After the black holes merge, the total entropy of the resulting black hole is
\bea
\mathcal{S}^{\prime} = \frac{1}{4G}[A^{\prime}
                       + 8\pi\alpha\chi(\mathscr{S}^{\prime})] \; .
\label{entropyafter}
\eea
Without knowing the specific details of the black holes in question, we cannot
say anything further about $\mathcal{S}$ and $\mathcal{S}^{\prime}$.  Let us
therefore consider for concreteness the simplest case -- the merging of two
Schwarzschild black holes in an asymptotically flat spacetime.  In this case
the cross sections of the horizons can only be spheres and therefore
$\chi(\mathscr{S}_{1})=\chi(\mathscr{S}_{2})=\chi(\mathscr{S}^{\prime})=2$.
This, together with the fact that the area of a Schwarzschild black hole is
related to its mass via $A=16\pi m^{2}$, implies that the entropies
$\mathcal{S}$ and $\mathcal{S}^{\prime}$ are given by
\bea
\mathcal{S} &=& \frac{4\pi}{G}[m_{1}^{2} + m_{2}^{2} + 2\alpha]
\label{s}\\
\mathcal{S}^{\prime} &=& \frac{4\pi}{G}[(m_{1} + m_{2} - \gamma)^{2} + \alpha] \; .
\label{sprime}
\eea
Here we included a small mass parameter $\gamma\geq0$ for the surface area of the
final black-hole state that corresponds to any mass that may be carried away by
gravitational radiation during merging.  The expressions (\ref{s}) and (\ref{sprime})
imply that $\mathcal{S}^{\prime}>\mathcal{S}$ iff
\bea
\alpha < 2m_{1}m_{2} - \gamma[2(m_{1} + m_{2}) - \gamma] \; .
\eea
Therefore the second law will be violated if $\alpha$ is greater than twice the
product of the masses of the black holes before merging minus a correction due to
gravitational radiation.


To summarize, we explored the role that the Gauss-Bonnet term can play in
four-dimensional general relativity.  In particular, we constructed a covariant
phase space for an isolated horizon and calculated the first law.  This led to
an expression for the entropy that is given by the area of the horizon plus a
correction term that is given by the Euler characteristic of the cross sections
of the horizon.  As was shown, the correction term can have some interesting
effects during the merging of two black holes, as the second law can be violated
for certain values of the GB parameter.  Therefore we have shown that the GB term
can have non-trivial physical effects in four dimensions, contrary to the common
assumption that the term is only significant in spacetimes with five or more
dimensions.

It would be interesting to investigate the quantum geometry of these ``topological''
isolated horizons by using the methods that were developed in \cite{abck,abk,domlew}.
Quantization of toroidal and higher-genus horizons in Einstein gravity with negative
cosmological constant has been recently considered by Kloster \emph{et al} \cite{kbd}.
Interestingly it was found that the toroidal horizon is the only one for which the
quantum entropy does not acquire any logarithmic corrections.


The author thanks Ivan Booth for discussions and for numerous suggestions that
improved the presentation of the manuscript, and Kirill Krasnov for correspondence.
The author also thanks the participants at BH$6$ and at CCGRRA$12$ for discussions
related to this work, especially Kristin Schleich and Don Witt.  The author is
supported by the Natural Sciences and Engineering Research Council of Canada.

\end{document}